\documentclass[12pt]{article}
\usepackage{epsfig}
\usepackage{amssymb}

\def\beq{\begin{equation}}
\def\eeq#1{\label{#1}\end{equation}}
\def\eeqn{\end{equation}}

\def\beqa{\begin{eqnarray}}
\def\eeqa#1{\label{#1}\end{eqnarray}}
\def\eeqan{\end{eqnarray}}

\let\bar=\overbar

\def\Dslash{\not{\hbox{\kern-4pt $D$}}}
\def\dslash{\not{\hbox{\kern-2pt $\del$}}}

\def\msb{{\bar{\ssstyle M \kern -1pt S}}}

\usepackage{fancyhdr,graphicx}
\def\Title#1{\begin{center} {\Large {\bf #1} } \end{center}}

\begin{document}

\Title{Quark nova inside supernova: Application to GRBs and XROs}

\bigskip\bigskip


\begin{raggedright}

{\it Jan Staff\index{Staff, J.}\\
Department of Physics and Astronomy\\
Louisiana State University\\
202 Nichols Hall, Tower Dr.
Baton Rouge, La\\
70803\\
USA\\
{\tt Email: jstaff@lsu.edu}}\\
\bigskip\bigskip
{\it Rachid Ouyed\index{Ouyed, R.}\\
Department of Physics and Astronomy\\
University of Calgary\\
SB 605
2500 University Drive NW 
Calgary, AB\\
T2N 1N4\\
Canada\\
{\tt Email: ouyed@phas.ucalgary.ca}}
\bigskip\bigskip
\end{raggedright}

\begin{abstract}
In this paper we consider a quark nova occurring inside an exploding star.
The quark nova ejecta will shock when interacting with the stellar envelope.
When this shock reaches the surface of the star, the energy is radiated
away. We suggest that this energy may be seen in X-rays, and show here that
this may explain some flares seen in the X-ray afterglow of long gamma ray
bursts (GRBs). A quark
nova inside an exploding star need not be followed by a GRB, or the GRB may
not be beamed towards us. However, the shock break-out is likely not beamed
and could be seen even in the absence of a GRB. We suggest that XRO 080109
is such an event in which a quark nova occurs inside an exploding star. No
GRB is formed, but the break out of the shock leads to the XRO.
\end{abstract}

\section{Introduction}

Quark stars (QSs) are hypothetical objects composed
of deconfined u, d, and s quarks. The strange quark matter (SQM) hypothesis
states that SQM may be the absolute ground state
of strong interacting matter rather than $^{56}Fe$ \cite{bodmer71,
witten84}, and therefore QSs would be stable objects. Even before the
formulation of the SQM hypothesis Itoh \cite{itoh70} discussed the possible
existence of QSs. 

QS masses and radii
are likely comparable to neutron stars (NSs), but slightly smaller radius due 
to their higher density. 
A possible formation scenario for QSs is that
the central density in a neutron star increases past a critical density at
which quarks deconfine. This can happen either due to spin down of a rapidly
rotating NS \cite{staff06}, or through accretion onto the NS. A QS
can then be formed in a quark nova (QN) \cite{ouyed02}, during which up
to $10^{53}$ erg can be released in an explosive event. The iron rich NS
crust will likely be blown away in the explosion. For a QN to occur
soon after the core collapse, either the NS magnetic field must be very high
($B\sim10^{14}-10^{15}$ G) and the NS rapidly rotating, or more likely
through accretion.
In \cite{ouyed09} the interaction of the iron rich QN ejecta and the
surrounding star was studied, and applied to long gamma ray bursts (GRBs). 
There it was suggested
that QN ejecta leaving the star through a funnel could lead to precursor
activity.


Long GRBs typically have a duration $T_{90}\sim 100$ seconds, after which the
gamma radiation ends. The X-ray telescope on board Swift is sometimes able
to focus onto a GRB at a timescale of around a 100 seconds. A generic X-ray
light curve \cite{zhang06} can be constructed based on XRT observations. It 
shows a sharp
drop in the X-ray light curve lasting for a few hundred seconds, followed by
a plateau phase extending to several times $10^4$ seconds, before a
``normal'' power law decay is seen. Overlayed on all this are one or more
flares, some of which can have a fluence comparable to that of the prompt
gamma emission itself yet typically they have a much lower energy. Other
bursts do not show the sharp drop and plateau phase, but rather a ``normal''
power law from early times. There may however be flares overlayed also on
this light curve.

The ``normal'' power law decay is what is expected from the external shock.
This power should be steeper than -0.75 for pre-jet-break and
steeper than -1.5 for post-jet-break light curves \cite{liang08}, yet not
steeper than -3 \cite{troja07}.
Hence the external shock alone has difficulty explaining the very sharp drop,
the flares, and the plateau phase, which is why other mechanisms involving
the inner engine must be sought in the cases where these features are seen.

In \cite{staff07} a three stage model for the inner engine of long GRBs was
suggested. Stage 1 is a NS formed by the collapse of the iron core in a SN.
Accretion or spin-down can make the NS explode in a QN leaving a QS behind,
which leads to stage 2 that is a jet launched from hyperaccretion onto a QS
giving the GRB prompt emission. When the QS has accreted sufficiently, it
collapses to a black hole (BH) whereby the third stage can be reached, which 
is accretion
onto a BH. This may launch another jet that can give rise to one big flare
(or several flare in rapid succession). If accretion ends before the QS
collapses to a BH, stage 3 will never be reached and instead spin down of 
the highly magnetized rapidly rotating QS
may power a secondary outflow giving rise to a flat segment in the X-ray
afterglow \cite{staff08}.

In this paper we build on the work in \cite{ouyed09} and explore the
breakout of the shock formed when the QN ejecta interacts with the stellar
envelope. We will use this to explain features seen in GRBs or XRO
080109. The expressions for describing the QN induced shock's propagation
through the stellar envelope is given in section \ref{theorysection}. In
section \ref{grbsection} we study possible observational consequences of the
QN that formed the QS leading to the GRB. Using the simple expressions
presented in \ref{theorysection} we show that some later flares in
GRBs may be formed by the ejecta launched from the QN interacting with the
stellar envelope. The QN ejecta are likely ejected isotropically, so even
when no GRB is seen these weaker and later flares can be seen. A QN does not
have to be followed by a GRB and in section \ref{xrosection} we suggest that 
XRO 080109 was a case where a QN occurred inside a SN, but no GRB was
produced. A summary of the three stage model and its features
is given in section \ref{summarysection}.

\section{QN ejected chunks interacting with stellar envelope}
\label{theorysection}

\begin{table}
\begin{tabular}{lll}
\hline\noalign{\smallskip}
$E_{\rm QN}$ & $10^{52}$ erg & Energy of QN going into chunks\\
$m_{\rm ejecta}$ & $10^{-5} M_\odot$ & mass of QN ejecta going into chunks\\
$M_{\rm env}$ & $1 M_\odot$ & Mass of the stellar envelope\\
$n_{\rm c}$ & 1000 & number of chunks \\
$R_{\rm sep}$ & $10^{10}$ cm & Distance at which chunks separate \\
$T_{\rm bb}$ & 10 keV & The blackbody temperature \\
$R_{\rm env}$ & $5\times10^{11}$ cm & Radius of the envelope \\
$\Gamma_{\rm spread}$ & $0.2 \Gamma_{\rm avg}$ & Spread in the chunks's
Lorentz factors\\
\hline
\end{tabular}
\caption{The free parameters used in our model, together with their typical
values and a brief explanation of their meaning.}
\label{freeparamtable}
\end{table}

The QN forming the QS ejects up to $0.01 M_\odot$ from the crust of
the NS. This QN ejecta breaks up into $n_{\rm c}$ chunks at a distance 
$R_{\rm sep}$. The
chunks are given Lorentz factors following a Normal distribution, with a standard
deviation of $\Gamma_{\rm spread}$, which is a free parameter. The average
Lorentz factor is 
\begin{equation}
\Gamma_{\rm avg}=\frac{E_{\rm QN}}{m_{\rm ejecta}c^2},
\end{equation}
where $E_{\rm QN}$ is the energy of the QN going into chunks, $m_{\rm
ejecta}$ is the mass of the QN ejecta going into chunks, and $c$ is the
speed of light.
In addition we impose a minimum Lorentz factor of 2 on the chunks. A larger
$\Gamma_{\rm spread}$ results in a longer duration of the chunks interacting
with the envelope. These
chunks undergo a shock and is heated up to a temperature
\begin{equation} 
T_{\rm c}= (\frac{\rho_{\rm env}}{\rho_{\rm Fe}})^2 \Gamma_c m_{\rm Fe} c^2
\end{equation} 
as they interact with the stellar envelope ($m_{\rm Fe} c^2=56$ GeV), where 
\begin{equation}
\rho_{\rm env}=M_{\rm env}/(4/3\,\pi r_{\rm env}^3)
\end{equation}
is the density in the envelope, and $\rho_{\rm Fe}\sim10{\rm g/cm}^3$ is 
the density of iron. The shock speed is given by 
\begin{equation} 
v_{\rm shock}=\sqrt{T_{\rm c}/\mu_{\rm env}}c, 
\end{equation} 
where $\mu_{\rm env}=1.2$ is the mean atomic mass in the envelope and
$T_{\rm c}$ is measured in GeV.
Each shock has an energy
\begin{equation} 
E_{\rm sh}=\pi \Delta r_{\rm c}^2 \sigma T_{\rm bb}^4,
\end{equation} 
where $\sigma=5.67\times10^{-5}$ is the Stefan Boltzmann constant, 
$T_{\rm bb}$ is the blackbody
temperature when the shock breaks out of the star, and
\begin{equation}
\Delta r_{\rm c}=\sqrt{\frac{4 R_{\rm sep}^2}{n_{\rm c}}}
\end{equation} 
is the chunk size at birth. This energy is released 
instantly as the shock reaches the
surface of the envelope at $t=r_{\rm env}/v_{\rm shock}$. Due to the
different shock velocities, the shocks will break free at different times.
Shocks breaking out at the same time will not necessarily be observed at the same
time, since the light travel time may be different depending on where on the
star the break out happened. However, the light crossing time of the star
(the maximum difference in light travel time) is much smaller than the
duration of the breakout of all the shocks, and so it is ignored in our
calculations.
We assume that all shocks have the same temperature $T_{\rm bb}$ when they
release their energy.

The QN explosion is likely to be more or less isotropic and
therefore the shocks from the chunks will be distributed
isotropically at all latitudes in the star. The fraction of the released 
luminosity ($L_{\rm emitted}$) that is observed is therefore 
\begin{equation}
L_{\rm obs}=\frac{L_{\rm emitted}}{4\pi d^2},
\end{equation}
where d is the luminosity distance to the event. Note that the shocks are
non-relativistic, so no relativistic beaming of the emitted radiation 
is present.

Table~\ref{freeparamtable} lists the parameters in our model.


\section{Application to long GRBs}
\label{grbsection}

In the three stage model for GRBs \cite{staff07}, the prompt gamma ray
emission is likely to start soon after the QN. We will for now assume that
the GRB jet is capable of making its way through the stellar envelope with a
high Lorentz factor ($\Gamma\sim100$). 
The QN ejecta is thus launched a short time before the
onset of the GRB.  As discussed above, the shock speed (and thus the time it
takes the shock to propagate through the envelope) is proportional to
$\rho_{\rm env}$. 

Here we consider the case of a rather thick envelope ($r>5\times10^{11}$
cm). We also envision that there is a lot of structure to this envelope, so
that some of the chunks can propagate unhindered through part of the
envelope, and only interact with matter at higher radii. Such a scenario may
be possible if there is a delay (of minutes to an hour) between the SN and
the QN. For simplicity we assume that there are two envelopes surrounding
the star, and the chunks can interact either with the first or the second.

\subsection{GRB 070110}

\begin{figure}
\includegraphics[width=0.7\textwidth]{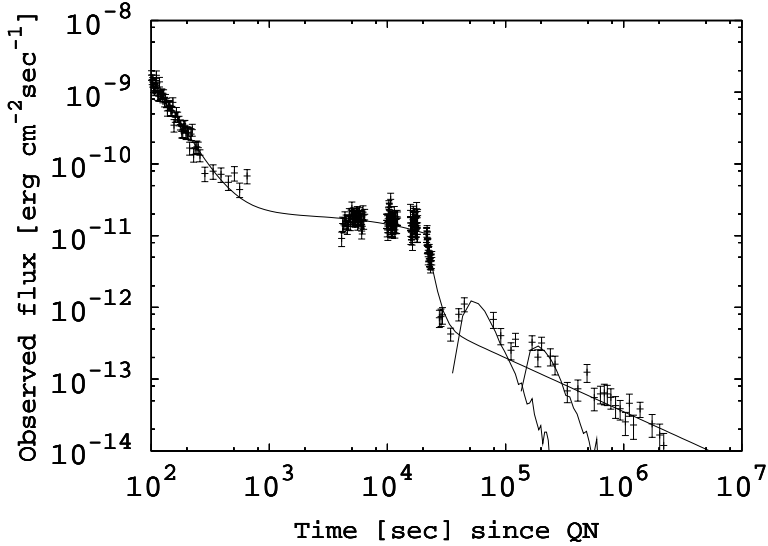}
\caption{The XRT data points for the X-ray afterglow of GRB 070110
\cite{evans07}. There
are several prominent features in this afterglow: a flat plateau lasting
until about 20000 seconds followed by a very sharp drop. After the sharp
drop follows a decaying light curve with two flares overlayed. In addition to
the data, we have plotted a power law plus the effect of a QS spinning down
from 2 ms with an initial magnetic field of $5\times10^{14}$ G
\cite{staff08}. At around t=20000
sec. the QS collapses to a black hole causing the abrupt drop in the light
curve down to the level given by the external shock. This we indicate with a
power law with power $-0.75$. Also shown are two flares due to the QN chunk
break out from the two layers of the stellar envelope, these flares are seen
at about 50000 sec. and 130000 sec. The time is the
time since the QN (essentially the time since the start of the GRB), and on
the vertical axis is shown the observed flux. $10^{-12} {\rm\,erg\, cm}^{-2}
{\rm \,s}^{-1}$
corresponds roughly to $5\times10^{46}$ erg/s assuming isotropic emission
and a redshift z=2.35.}
\label{070110fig}
\end{figure}

The X-ray afterglow light curve for this burst is quite remarkable. From about
4000 seconds to about 20000 seconds after burst trigger, the light curve
looks flat. Following that, there is a very steep decay (a slope of power
$\alpha>7$)\cite{troja07}. Following the flat segment and this steep
decay, there are two flares overlayed on a power law decay. In \cite{staff07}
it was suggested that the flat segment is powered by spin down of
a QS. The flat segment abruptly ends as the QS collapses to a BH, and the
light curve drops sharply to the level given by the external shock. In this
section we suggest that the flares seen in the light curve following the
flat segment and the sharp drop is due to the QN chunks breaking out of the
envelopes (see Fig.~\ref{070110fig}).

The parameters we have used to obtain this plot is:
$m_{\rm ejecta}=0.8\times10^{-4}M_\odot$, $E_{\rm QN}=10^{52}$ erg, $M_{\rm
env}=0.5 M_\odot$, $r_{\rm env, 1}=9\times10^{11}$ cm, $r_{\rm env,
2}=11\times10^{11}$ cm, $R_{\rm
sep}=3\times10^{10}$ cm, $T_{\rm bb}=25$ keV, $\Gamma_{\rm
spread}=0.2\times \Gamma_{\rm avg}$. We note that the envelope is at a
fairly large radius, indicating that the supernova exploded some time before
(some minutes depending on the size of the exploding star and the velocity
of the ejecta) the QN and the GRB occurred. The redshift for this burst is 
$z=2.3$, so although the
$T_{\rm bb}$ seems large for X-ray observations, the radiation will be
redshifted from 25 keV to 7.6 keV. This is in the range
of the XRT.

The total energy released as X-rays in the two flares is $7.2\times10^{50}$
erg for the first flare and $5.9\times10^{50}$ erg for the second flare.
This is more than $10\%$ of the total energy ($10^{52}$ erg) released 
in the QN. Hence the efficiency in converting energy into X-rays must be
fairly high.
Because the flares are so broad due to the large envelope size the
observed flux is not very high. 

We have matched three power laws, the spin down of a rapidly rotating quark
star, and two flares due to QN chunk breakout to
the data (see Fig.~\ref{070110fig}). First the initial decay lasting until 
500 seconds with a power
$-2.8$ which may be due to curvature effect. Following this is the spin down
light curve of a rapidly rotating highly magnetized QS
lasting until $t=20000$ seconds. At t=20000 seconds the QS collapses to a
BH ending the flat segment. The light curve drops abruptly (we use a power
law with power $-9.3$) to the level given by the external shock. Following
the break is another power law with power $-0.75$ overlayed with the two QN
chunk flares seen around $10^5$ seconds.

\subsection{GRB 081007}

\begin{figure} 
\includegraphics[width=0.7\textwidth]{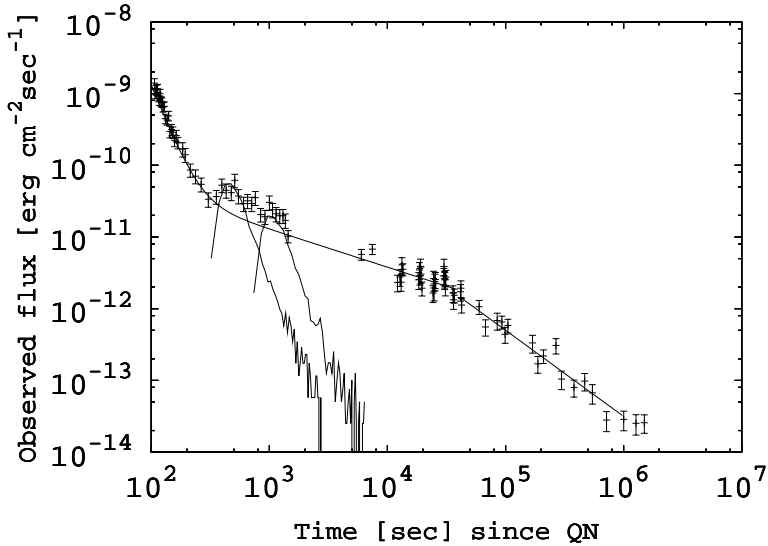}
\caption{The XRT data points for the X-ray afterglow of GRB 081007
\cite{evans07}. Two
flares due to QN chunk break out is shown after 500 seconds and 1000
seconds. In addition, we plot three power laws, until 250 seconds the power
is $-4.1$, then until 40000 seconds it is $-0.53$, followed by a third power
law with power $-1.2$. The break that we found at 40000 seconds may be a
jet-break \cite{rhoads99}, but the power following the break is fairly
shallow indicating that this is still pre-jet break. Instead, we
suggest that there is a period of refreshed shocks ending at 40000 seconds
followed by the pre-jet-break decay.}
\label{081007fig}
\end{figure}

GRB 081007 shows a light curve with a fairly steep decay until $t\sim300$
seconds. After that there is a small flare around 500 seconds and another at
around 1000 seconds. We assume that these flares are due to the QN chunks
breaking out and plot the light curve in Fig.~\ref{081007fig}.

The parameters used to obtain this plot is:
$m_{\rm ejecta}=1\times10^{-4}M_\odot$, $E_{\rm QN}=10^{52}$ erg, $M_{\rm
env}=5 M_\odot$, $r_{\rm env, 1}=6\times10^{11}$ cm, $r_{\rm env,
2}=7.4\times10^{11}$ cm, $R_{\rm
sep}=1\times10^{10}$ cm, $T_{\rm bb}=10$ keV, $\Gamma_{\rm
spread}=0.2\times \Gamma_{\rm avg}$.

The redshift for this burst is $z=0.5$,
so the observed peak will be seen at 6.7 keV instead of the 
emitted 10 keV. This is within the XRT range.

The total energy released in X-rays in the first flare is $8.8\times10^{48}$
erg and in the second flare it is $7.2\times10^{48}$ erg. In this case a
much lower fraction of the total QN energy (compared to GRB 070110) has been
released as X-rays in these two flares. Because of the smaller envelope
radius, the observed flux in these flares is higher than in the GRB 070110 case.

We have also matched three power laws to the data. The first power law has a
power of $-4.1$ until t=250, this we suggest is the curvature emission
marking the end of the QS jet. Then another with power $-0.53$ until
$t=40000$ at which time a break is found. This break may be a jet break
\cite{rhoads99} or the end of a flatter segment due to refreshed shocks
\cite{reesmeszaros98}. Following the break is another power law function
with power $-1.2$. Since the light curve following the break is fairly
shallow, this may indicate that the jet break occurs later and that this
break is due to the end of a period of refreshed shocks (for a
discussion on the pre-jet break power see for instance \cite{liang08}).

\subsection{Flares in GRB afterglows}

The sharp drop between 100 and 1000 seconds is often thought to
be ``curvature'' radiation \cite{kumarpanaitescu00}, that is the emission
emitted outside the $1/\Gamma$ cone. This would mark the end of the jet
that created the prompt emission (the QS jet in stage 2). Flares occurring
before this sharp drop could therefore be due to the same mechanism as that
which created the prompt emission, with the difference that the radiation is
seen by XRT rather than BAT. 

As mentioned in the introduction, if stage 3 is reached the BH jet may lead
to one big flare. There may be a delay between the end of the QS jet and the
launching of the BH jet, in which case this flare would be seen after the
beginning of the sharp drop. It is also
possible that there are no delay between the QS jet and the BH jet, in which
case this big flare from the BH jet will occur shortly before a possible
sharp drop\footnote{Accretion onto a QS will only power a jet for a certain
range in accretion rates ($\dot{m}$). When $\dot{m}$ is higher than what the 
magnetic field
can channel to the polar cap then no jet is launched. This accretion rate is
found by requiring that the Alfv{\'e}n radius be larger than the star,
$r_A=\big(\frac{B^4r^{12}}{2GM\dot{m}^2}\big)^{1/7}$. For a star with
$10^{15}$ G magnetic field, this critical $\dot{m}$ is about 
$10^{-3}M_\odot/sec$. On the contrary,
$\dot{m}$ must be sufficiently high (higher than about $10^{-6}M_\odot/sec$) 
that it can heat the QS surface above
$7.7$ MeV in order for the QS to cool by photon emission (which is how the
jet is launched). For accretion rates lower than this, no jet will be
launched. If the accretion rate evolves over time, $\dot{m}$ may initially
be in the range that can launch a jet, but later evolve out of this range.}.

The maximum mass that can be accreted onto the QS is assumed to be of the
order $0.1 M_\odot$, whereas the maximum mass that can be accreted onto the
BH is all the mass in the star that is not expelled in the explosion, 
probably up to $\sim
10M_\odot$ (although we expect much smaller mass accreted to be the norm). 
The accretion rate is likely very high
($\dot{m}\sim1M_\odot/sec$), leading to a short duration. The energies
involved can potentially be substantially higher than in the prompt phase.

In this paper we have suggested a third mechanism to form flares, when the
QN ejecta shocks the stellar envelope and these shocks break out. The time
for the occurrence of these flares depends on the thickness of the envelope,
which is dependent on the initial star as well as the time delay between the
core collapse and the QN. These flares can therefore occur at almost any
time, both early and late. However, were they to occur before the sharp
drop off following the GRB jet, they are unlikely to be seen as they would
not be sufficiently bright to outshine the jet.

In bursts where the ``normal'' power law is seen from early times, we
suggest that the external shock is sufficiently bright that it outshines
possible plateau phases. Only the top of the brightest flares can be seen.
For a sufficiently fast drop off of the light curve, the QN chunk break out 
may be observed at later times in this scenario as well.

%
%

\section{Application to XRO 080109}
\label{xrosection}

XRO 080109 is an X-ray outburst associated with SN 2008D which is a type
1b/c SN. The XRO lasted about 500 seconds, with a peak about 65 seconds
after trigger \cite{soderberg08}. The peak X-ray luminosity is about
$10^{44}$ erg/sec, decaying down to a few times $10^{42}$ erg/sec after
about 700 seconds \cite{xu08}. No GRB was seen in connection with this
event.
either means that there really was no GRB in this case, or that the GRB was
beamed away from us. Radio observations and other indications
\cite{soderberg08} argues in favor of no GRB. We adopt this view here.

We propose that in XRO 080109 the iron core of the star collapsed to a
(proto-) neutron star. The mechanism that explodes a supernova acted.
Shortly
thereafter, the compact core collapsed to a QS in a QN. A QS was therefore
formed, but that does not necessarily mean that a GRB was produced. There
will be no GRB (in our model) without a hyperaccretion disk forming around 
the QS, when no jet is formed or the jet cannot escape the stellar envelope. 
A hyperaccretion disk may be
more likely to form in a system whose progenitor was rapidly rotating. 

\begin{figure}
\includegraphics[width=0.7\textwidth]{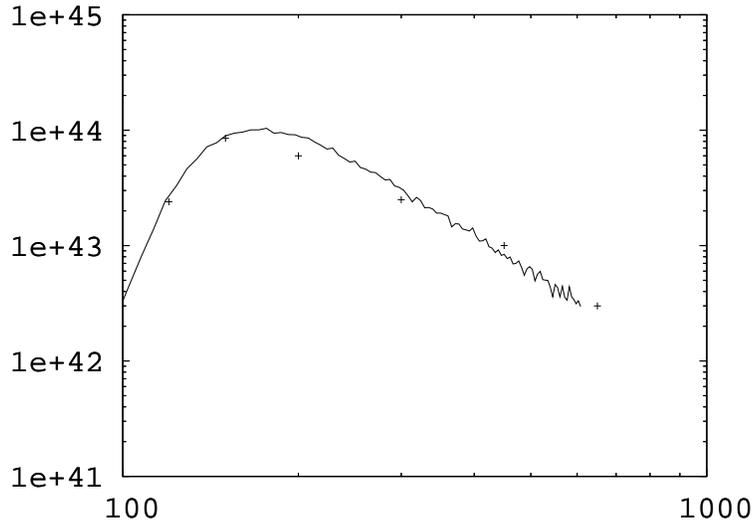}
\caption{The model light curve for XRO 080109. The time scale on the
horizontal axis is the time since the explosion.}
\label{xro080109model}
\end{figure}

In Fig.~\ref{xro080109model} we show the light curve for our model XRO
080109. The parameters used to obtain this was $M_{\rm envelope}=7M_\odot$, 
$M_{\rm ejecta}=5\times10^{-4}M_\odot$, $E_{\rm QN}=10^{52}$ erg, $T_{\rm
bb}=1.7$ keV, $r_{\rm envelope}=5\times10^{11}$ cm, $\Gamma_{\rm
spread}=0.45\Gamma_{\rm avg}$. An envelope mass of
$7M_\odot$ was found in \cite{mazzali08}. We note that $r_{\rm
envelope}\sim5\times10^{11}$ cm is roughly consistent with the estimate in
\cite{soderberg08} that the shock breakout occurs at
$r>7\times10^{11}$ cm. They speculate that this is shock breakout from a
wind surrounding the exploding star. In our model, the shock breakout could
be simply from the stellar envelope, provided that there was a delay between
the core collapse and the QN, during which the envelope had time to expand.

In \cite{li08} it is shown that the spectra of XRO 080109 can be fitted with two
blackbody components, or one power law. The two blackbodies could be due to
shock breakout of the SN shock and the QN shock. We note, however, that 
\cite{li08} found a photospheric radius much smaller than a solar radius, 
something that needs to be further explored in our model.

\section{Summary}
\label{summarysection}

In this paper we have explored the break-out of the shock created by a QN
ejecta when it interacts with the stellar envelope surrounding the QN. The
QN is assumed to occur inside an exploding star. In
the event of a non-uniform envelope the resulting light curve from this
shock break-out can become complicated. Here we simplified the
non-uniformity to assuming that there are two envelopes, one outside the
other, and that a fraction of the QN ejecta interacts with the inner
envelope and the rest with the outer. Based on this we find that two flares
can result from the shock break-out. We suggest that these flares can
explain some of the flares seen in GRB afterglows. The shock break-out is
likely not beamed, whereas GRBs are. A GRB
not beamed towards us will not be seen, while the QN shock break-out
can still be seen. A QN can also occur inside a SN without forming a GRB at
all, while the shock break out can be seen. We suggest
that this is the case in XRO 080109.

We here summarizes the three stage model, and emphasizes the new aspects
proposed in this paper:

\begin{itemize}

\item
The iron core in an initially massive, rapidly rotating star collapses,
creating a (proto-)neutron star. This is the first stage in the model.

\item
Through spin down of the NS or accretion onto the NS (or both), the central 
density in the QS reaches a critical value at which a QN occurs, creating a
highly magnetized QS.

\item
A hyperaccretion disk surrounding the QS accretes onto the QS, launching an
ultrarelativistic jet in which internal shocks produces the GRB. This is the
second stage in the model.

\item
The jet launching continues until the accretion is no longer able to heat
the QS above 7.7 MeV, until accretion is no longer channeled to the polar
cap (because the QS magnetic field drops or the accretion rate increases), or
until the accretion halts (because there is no disk left).
This leads to a sharp drop off due to curvature radiation frequently seen in 
the X-ray afterglow. If the external shock has already formed and is strong
before the sharp drop off, the sharp drop off may not be seen.

\item
Flares seen before the sharp drop off may be caused by the same mechanism
that created the GRB itself (internal shocks), only that the radiation is 
softer.

\item
If sufficient mass is accreted onto the QS, it will collapse to a BH and 
Stage 3 is reached which is continued accretion onto the BH. This can 
launch another jet, leading 
to a major X-ray flare. This flare will likely occur after the sharp drop 
off started, but not necessarily before it ended (it may also occur after
the end of the sharp drop off).

\item
We assume that a maximum of $0.1 M_\odot$ can be accreted onto the polar cap.
Usually accretion rates of the order $\dot{m} \gtrsim 10^{-4} M_\odot/sec$ is 
needed to explain the observed GRB power, indicating a total duration of the
QS phase $t\lesssim 1000 sec$. However, if the accretion rate drops during
stage 2, this duration may be extended.

\item
In the event that the QS survived the accretion stage, stage 3 will not be
reached. Because of the accretion, this QS is likely rapidly rotating, and
we assume it to be highly magnetized. Hence rapid spin down due to magnetic
braking will occur. This can lead to an observed flat segment in the
afterglow, if the spin down power outshines the external shock. The QS may
reach an unstable state during spin down, leading it to collapse to a BH.
The observed light curve will then drop sharply down to the level given by
the external shock.

\item There may therefore be two separate outflows leading to two different
light curves seen for instance as the X-ray and
optical afterglow. One is created by spin down and the other by the
external shock.

\item
In this paper we suggested that the shocks created by the chunks ejected in
the QN can create observable flaring when it breaks out. In some cases
this may be observed, in others it may be dwarfed by the external shock.
The time at which these shocks break out depends largely on the size of the
exploding star's envelope, which in turn depends on the delay between the
collapse of the iron core and the QN. Here we have showed how two flares 
may result if the exploding star's envelope is not uniform.

\item
A flatter than expected light curve is also possible with the refreshed
shock mechanism, in which slower part of the QS jet catches
up with the external shock at later time and ``refreshes'' it leading to a
shallower light curve. At the end of the refreshing, a break in the light
curve is observed as the light curve steepens when no more material can
refresh the external shock.

\item 
A jet break can be observed in the light curve when
$1/\Gamma>\theta_{\rm jet}$ ($\Gamma$ being the 
Lorentz factor of the jet and $\theta_{\rm jet}$ being the opening angle 
of the jet itself) at which point the light curve steepens. Since there can
be two separate outflows leading to the optical and X-ray afterglow, this
may explain why the breaks (thought to be jet-breaks) are not achromatic. 

\item
The QS jet, BH jet, and QS spin down outflow are all assumed to be beamed in
two bipolar jets. The QN ejecta on the other hand is assumed to not be 
beamed, and it could therefore still be possible to see this even when the
GRB is not beamed towards us. A QN may also occur inside an exploding star
without forming a GRB (this happens if no hyperaccretion disk forms around
the QS). However, the chunk break-out may still be seen. We have here
suggested that XRO 080109 is such an event in which a QN occurred but no GRB
was formed. 

\end{itemize}

\bigskip
The authors would like to thank the organizers of CSQCDII for a very
interesting and fruitful meeting.
This work has been supported, in part, by grants AST-0708551, PHY-0653369,
and PHY-0326311 from the U.S. National Science Foundation and, in part, by
grant NNX07AG84G from NASA’s ATP program.

\end{document}